\documentclass[conference]{IEEEtran}
\usepackage{amsmath,graphicx}
\graphicspath{{images/}}
\usepackage{amsfonts}

\usepackage{bm}
\usepackage{multirow}

\ifCLASSINFOpdf
\else
\fi
\graphicspath{{Images/}}

\DeclareMathOperator*{\maxi}{max}

\IEEEoverridecommandlockouts

\begin{document}
\title{Rumour Detection via News Propagation Dynamics and User Representation Learning }

\author{
\IEEEauthorblockN{
Tien Huu Do$^{\star \dagger}$,~~Xiao Luo$^{\star \diamond \ast}$\thanks{$^{\ast}$The work was done when Xiao Luo was a student at Vrije Universiteit Brussel, Belgium},~~Duc Minh Nguyen$^{\star \dagger}$,~~Nikos Deligiannis$^{\star \dagger}$
}
\IEEEauthorblockA{
$^{\star}$ Vrije Universiteit Brussel, Pleinlaan 2, B-1050 Brussels, Belgium \\
$^{\dagger}$ imec, Kapeldreef 75, B-3001 Leuven, Belgium \\
$^{\diamond}$ CRRC Zhuzhou Institute Co., Ltd., 412001 Shidai Road, Zhuzhou, Hunan\\
\{thdo, mdnguyen, ndeligia\}@etrovub.be, luoxiao5@csrzic.com}
}

%\IEEEauthorblockA{
%Vrije Universiteit Brussel, Pleinlaan 2, B-1050 Brussels, Belgium \\
%imec, Kapeldreef 75, B-3001 Leuven, Belgium \\ 
%}


%Telephone: (800) 555--1212, Fax: (888) 555--1212}


\maketitle

\begin{abstract}
Rumours have existed for a long time and have been known for serious consequences. 
The rapid growth of social media platforms has multiplied the negative impact of rumours; it thus becomes  important to early detect them. 
Many methods have been introduced to detect rumours using the content or the social context of news. 
However, most existing methods ignore or do not explore effectively 
the propagation pattern of  news in social media, including the sequence of interactions of  social media users with  news across time. 
In this work, we propose a novel method for rumour detection based on deep learning. 
Our method leverages the propagation process of the news by learning the users' representation and the temporal interrelation of users' responses. 
Experiments conducted on Twitter and Weibo datasets demonstrate the state-of-the-art performance of the proposed method.
\end{abstract}
%
%\begin{keywords}
%Twitter user geolocation, multiview learning, deep learning, feature learning.
%\end{keywords}
%
%
%
%----------------------------------------------------%
\section{Introduction}

Rumours are items of unverified circulating information~\cite{zubiaga2018detection}, which have been known for serious consequences. 
The growth of social media platforms creates fertile ground for rumours, thereby rendering rumour detection of great significance. 
However, detecting rumours is a challenging task; studies have reported that humans are not good at identifying rumours~\cite{shu2017fake}. 
On the other hand, researchers have studied rumours from different points of view. 
There exist two prominent approaches for rumour detection: the content-based and social-context-based approaches. 
In the content-based approach, rumours are detected based on the content of news and  prior knowledge extracted from vast data sources~\cite{ciampaglia2015computational} or the writing style of the news~\cite{rubin2015truth}. 
Alternatively, the social-context-based  approach exploits the social engagements of social media users, e.g.,~replies on Twitter~\cite{duc2019naccle}. 
Using this approach, the massive quantity of user opinion can be aggregated, revealing the credibility level of the news~\cite{jin2016news}. Furthermore, the social-context-based methods can uncover the hidden temporal propagation pattern of the news~\cite{wu2015false}. As such, the social-context-based approach has recently become popular thanks to its good performance and the availability of additional information~\cite{zubiaga2018detection}.

\begin{figure}[htp]
\centering
\includegraphics[width=\linewidth]{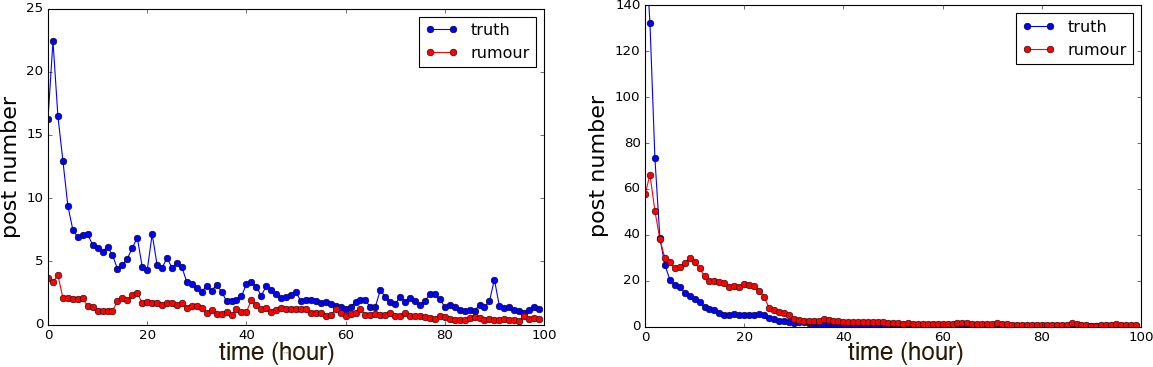}
\caption{Responses of social media users toward news on the Twitter (left) and Weibo (right) datasets~\cite{castillo2011information,ma2016detecting,ma2015detect}. For each one-hour interval, we calculate the average number of social posts associated to all news articles. 
The blue and red lines show the average number of posts for genuine news and rumours, respectively.
}
\label{fig:post_pattern}
\end{figure}

In this work, we address the problem of rumour detection on social media using social context information. 
We consider it a binary classification problem with two classes, i.e.,~\textit{non-rumour} and \textit{rumour}. 
By analyzing existing datasets, i.e., the Twitter and Weibo datasets~\cite{castillo2011information,ma2016detecting,ma2015detect}, we observed some peculiarities in the propagation process of news through social media users. 
Firstly, there is a difference in the numbers of posts towards rumours and genuine news across time instances, as illustrated in Fig.~\ref{fig:post_pattern}. 
Secondly, some users are more vulnerable to misleading information than others. 
As a result, these users tend to be involved in the spreading of many rumours in social media. 
Inspired by these observations, we aim to detect rumours by recognizing the peculiarities of the propagation process of the news.  
To this end, we design a novel propagation-driven model based on recurrent neural networks (RNNs) for rumour detection, which we name  \textit{Dual RNN for Rumour Detection} (DRRD). 

Our contributions in this paper are:
(\textit{i}) we propose the DRRD model, which can effectively learn the propagation pattern of news via its social engagements. We conjecture that the propagation pattern is an important factor to detect rumours. Furthermore, 
(\textit{ii}) we design a novel \textit{padding-and-scaling} procedure to improve the input features of the proposed model leveraging our observations; 
 (\textit{iii}) we propose a novel user representation learning technique exploiting the historical interactions of social media users across multiple news articles; and (\textit{iv})~we perform a series of experiments on two benchmark datasets and show that our model outperforms the existing methods in detecting rumours. 

The rest of this paper is organized as follows. In Section~\ref{sec:related_work} we review related studies. The details of our method are given in Section~\ref{sec:method}, and the experimental study is presented in Section~\ref{sec:experiments}. Section~\ref{sec:conclusion} concludes our work.

%\begin{figure}[t]
%\centering
%\includegraphics[width=\linewidth]{user_histogram} \\
%
%\caption{Number of users vs number of events.}
%\label{fig:post_pattern}
%\end{figure}
%\vspace{-2mm}

\section{Related Work}\label{sec:related_work}

The content-based rumour detection approach considers the textual content of news. Methods following this approach can be further divided into \textit{knowledge-based} and \textit{style-based}. 
Knowledge-based methods often rely on domain experts to perform rumour detection, and thus, require a huge amount of laborious effort. 
Moreover, human experts cannot keep up with the enormous volume of online information.
Therefore, computational knowledge-based methods have been introduced, including the~\textit{key fact extraction}~\cite{magdy2010web} and the~\textit{knowledge graph}~\cite{ciampaglia2015computational} methods. 
On the other hand, the style-based methods leverage the language peculiarities of the news to detect rumours by using natural language processing (NLP) features, such as \textit{lexical}, \textit{part-of-speech}, \textit{linguistic inquiry and word count} (LIWC) or \textit{deep syntax} features~\cite{mukherjee2013fake,rubin2015truth,rashkin2017truth,potthast2017stylometric}. 
Style-based methods do not require additional data; however, their performance is limited as misleading information is often manipulated meticulously, making it difficult to detect deceptive writing styles.
There also exist content-based methods that exploit news' creator profiles, partisan information or enclosed media. These methods often employ deep learning models, leveraging their advantage of fusing high-level features~\cite{yang2018ti,wang2017liar,zhang2018fake}. 
Although different types of information about news are integrated in these models, the propagation pattern of the news is ignored. 
In contrast, our method is not based on the news content; instead, we focus on the propagation process of the news and the interactions of social media users.

Alternative methods rely on the reactions of social media users towards news. These methods can be subcategorized into \textit{stance-based} and \textit{propagation-based} ones. In the \text{stance-based} methods, the viewpoints of relevant posts are taken into account to assess the veracity of the news. 
This idea has been realized in~\cite{jin2016news},~\cite{tacchini2017some} using \textit{label propagation} and \textit{boolean label crowdsourcing} (BLC), respectively. 
Alternatively, a number of studies have proposed to leverage the propagation process  by means of retweet trees~\cite{castillo2011information}, temporal interrelation~\cite{ma2015detect}, conditional random fields~\cite{zubiaga2016learning}, or a hierarchical propagation model~\cite{jin2014news}. 
Recently, many studies have applied deep learning for debunking rumours based on the propagation pattern by using recurrent neural networks (RNNs)~\cite{ma2016detecting,chen2017call,kochkina2018all}, convolutional neural networks (CNNs)~\cite{yu2017convolutional,qian2018neural} and combined CNN-RNN models~\cite{liu2018earlydetect}. 
In~\cite{ruchansky2017csi}, a deep neural network model was proposed for fake news classification. While the model is able to effectively capture the temporal propagation pattern of the news, its capacity to generalize to unseen users is restricted because of the singular-value-decomposition (SVD) based approach deployed to learn the user feature. 
Motivated by~\cite{ruchansky2017csi}, we design a novel model capable of learning the propagation pattern from multiple perspectives. 
Furthermore, we devise a special \textit{padding-and-scaling} procedure to support the learning of the propagation pattern. 
To overcome the limitation of the SVD-based approach in~\cite{ruchansky2017csi}, we propose using a~\textit{doc2vec}~\cite{le2014distributed} model to learn the users' representation, which is generalizable to unseen users and less computationally expensive to calculate.

\section{The proposed rumour detection method}\label{sec:method}

\begin{figure}[t]
    \centering
    \includegraphics[width=0.96\linewidth]{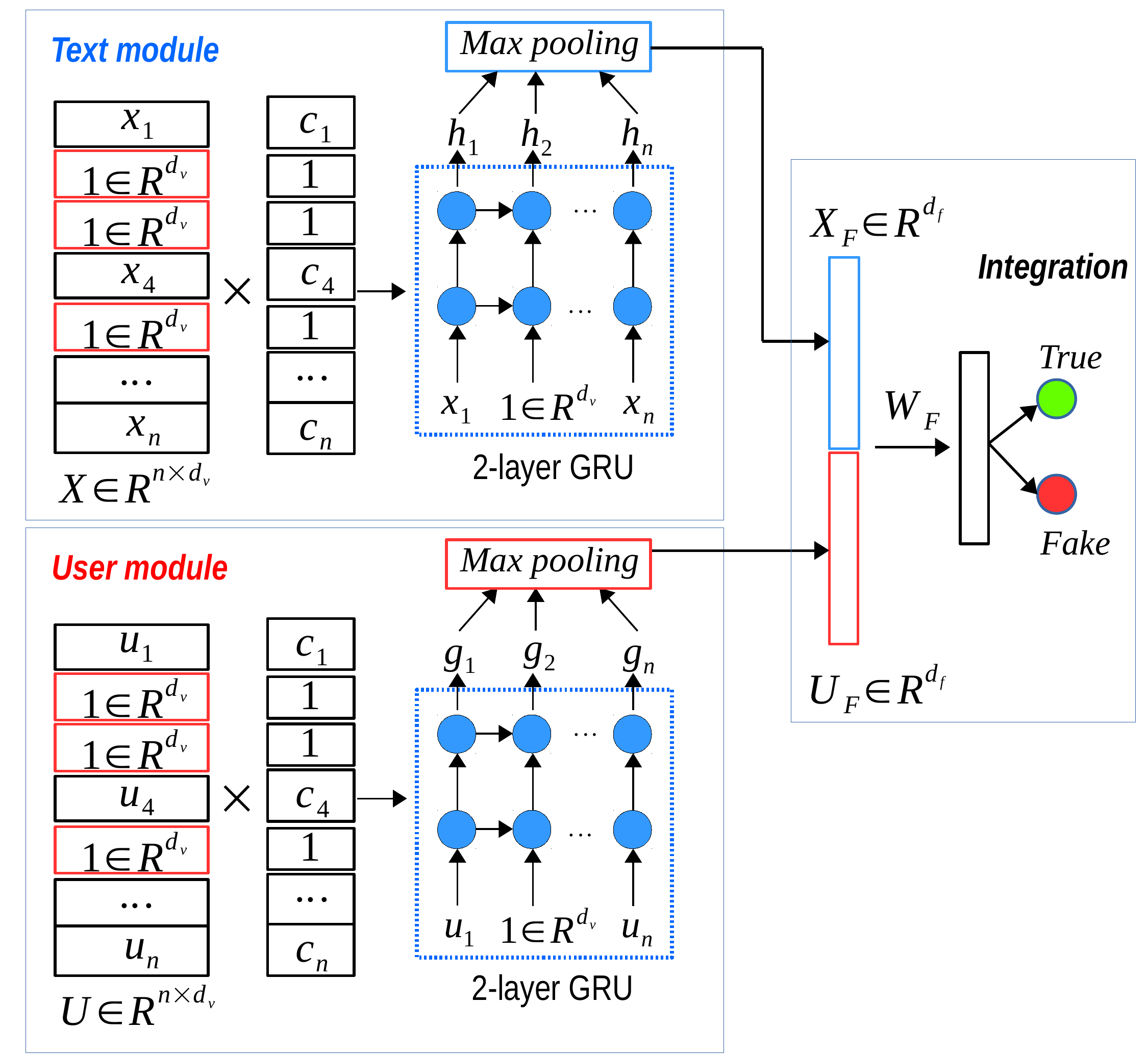}
    \caption{The architecture of the proposed DRRD model.}
    \label{fig:drrd_struc}
\end{figure}

\subsection{Problem Formulation and Notation}

We address the rumour detection problem using social context information. 
Let us assume that a news article reports a unique event and let $E = \{e^{(i)}\}^N_{i=1}$ be the set of such events.  An event $e^{(i)}$ has multiple social engagements, which refer to posts on social networks created by users that share or like the corresponding news article. 
Let~$S^{(i)}$ define the set of social engagements concerning the event~$e^{(i)}$, then $S^{(i)} = \{ (p_j, u_j, t_j)\}_1^M$, where $p_j$ represents the social post, $u_j$ is the user who makes the post, and $t_j$ is the corresponding timestamp. 
Let $L = \{ 0,1 \}^N$ be the binary label set of the events. 
Our goal is to establish a mathematical model $ \mathcal{F} $ predicting the probability for an event $e^{(i)}$ to be a rumour given its social engagements $S^{(i)}$, that is, $P(e^{(i)} = 1 | S^{(i)}) = \mathcal{F}(S^{(i)}) $

Concerning the early detection of rumours, we consider a set of social engagements within a deadline $T$. Let $S^{(i)}_T = \{ (p_j, u_j, t_j) \ | \ t_j < T \}_1^{M'}$ define the set of social engagements established before the deadline $T$, then the rumour probability of the event $e^{(i)}$ within $T$ is 
$ P(e^{(i)}=1 | S^{(i)}_T) = \mathcal{F}(S^{(i)}_T)$. 

\subsection{Data Partitioning Strategy}
In order to exploit the propagation pattern of news on social media, the relevant social posts have to be organized following a chronological order, i.e., by means of partitioning. 
For example,~\cite{yu2017convolutional} divided the posts into partitions of different time intervals such that the numbers of posts in the intervals are equal. 
However, we argue that the partitioning technique in~\cite{yu2017convolutional} ignores the intrinsic variation in the number of posts across the propagation process of the news, as indicated in Fig.~\ref{fig:post_pattern}. 
Therefore, we follow a natural way of partitioning by grouping posts by hour~\cite{ma2015detect, ruchansky2017csi}. 
Specifically, the timestamp of the earliest post concerning an event indicates the first appearance of the event. 
Moreover, the difference in hour(s) of a relevant post and the earliest post defines the \textit{hour index} of the post. The posts of an event with the same \textit{hour index} are then put into the same partition. 
An event is thereby represented by a sequence of hour partitions. 
We introduce a special \textit{padding-and-scaling} technique to promote the variation of posts in partitions,  presented in the following section. 

\subsection{Model Intuition and Structure}
Our model, which is depicted in Fig.~\ref{fig:drrd_struc}, is based on recurrent neural networks. It consists of three modules, namely, the~\textit{Text},~\textit{User} and \textit{Integration} modules.

\subsubsection{The Text Module}
In~\cite{ma2016detecting}, it was shown that the frequency of question words in  rumour posts is much higher than in non-rumour posts in certain time windows. 
Furthermore, as indicated in Fig.~\ref{fig:post_pattern}, there exists a difference in the number of social posts regarding rumours and true news. 
The text module is designed to capture these patterns.

Firstly, using the corpus of  social posts associated with the events
in the training set we train a~\textit{doc2vec} model~\cite{le2014distributed}, which has been proven useful in many NLP-related tasks~\cite{do2017multiview, duc2019naccle}. Using the trained~\textit{doc2vec} model, we obtain an embedding with $d_v$ dimensions for each social post. 
Subsequently, the embeddings of the posts in the same hour partition are averaged element-wise, constructing the representation of the  partition. 
We employ identity vectors, i.e., vectors with all~$1$ entries to represent partitions that contain no posts. 
An event is, therefore, represented by a matrix $X \in \mathbb{R}^{n \times d_v}$, where $n$ is the number of hours partitions. 
Each partition embedding is then scaled by a logarithmic coefficient defined by 
\begin{equation} \label{eq:log}
        c_k = \log(m_k + 1) + 1 \ ,
\end{equation}
where $m_k$ is the number of posts of the $k$-th partition. 
%We take the logarithm of $n_k + 1$ to avoid an invalid logarithm. We further add 1 in equation~(\ref{eq:log}) to avoid a zero coefficient in case $n_k = 0$. 
The purpose of this scaling is to capture the variation of the number of posts across partitions. 
Moreover, the logarithm is used to smoothen the coefficients as the values of $m_k$ may vary significantly across the partitions; for instance, the number of posts within an hour in the Weibo dataset ranges from $1$ to $24192$ posts. 

The padded and scaled representation is then passed to a two-layer RNN~\cite{lipton2015critical}. 
We choose the \textit{gated recurrent units} (GRUs)~architecture \cite{gru2014learning}  as it is easier to train compared to the \textit{long short-term memory} counterpart (LSTMs)~\cite{lstm1997long}, which was deployed in~\cite{ruchansky2017csi}. 
We, then, track the outputs $h_k\in \mathbb{R}^{d_f}$ of the RNN for all time steps~$k=1,\dots,n$, with~$d_f$ denoting the dimension of the output vector,  and apply \textit{max-pooling-over-time} to obtain  the  output feature vector~$X_F \in \mathbb{R}^{d_f}$ of the \textit{Text} module. Namely, the $l$-th element of the output feature vector is calculated as
\begin{equation}\label{eq:max_pooling}
X_{F_l} = \maxi_{k}  \{ h_{k,l} \}_{k=1}^n 
\end{equation}

\subsubsection{The User Module}
The user module is designed to capture the involvement of social media users in the propagation  of news. 
In~\cite{ruchansky2017csi}, it was shown that suspicious users tend to present a group behaviour, namely, most suspicious users are often involved in the rumours. 
To leverage this behaviour,~\cite{ruchansky2017csi} established a user adjacency matrix, which was  factorized using the SVD to obtain a representation for all users. 
However, the method is computationally expensive, especially for a large number of users, and non-scalable, since the adjacency matrix and, in turn, the SVD need to be re-calculated for every new user. 

Unlike~\cite{ruchansky2017csi}, we do not focus on the group behaviour but the sequence of user interactions with events across time. Specifically, we encode each user as a short document whose words are the names of the events the user interacts with. For instance, if the user $u_\ell$ tweets about the events $e^{(0)}$, $e^{(1)}, e^{(5)}$ and $e^{(10)}$, we use the document of the names~$\big[ e^{(0)}, e^{(1)}, e^{(5)}, e^{(10)} \big] $ to represent $u_\ell$. The resulting document is then
used to learn the user representation by means of the \textit{doc2vec}~\cite{doc2014distributed}
model. Per hour partition, the embeddings of users  are averaged and scaled [using~\eqref{eq:log}], similarly to the operations in the text module.
The resulting embedding per partition is passed to a two-layer RNN network; then,  \textit{max-pooling-over-time} is applied yielding the output $U_F \in \mathbb{R}^{d_f}$
of this module.

It is worth noting that, as shown in Table~\ref{tab:datasets}, a user  makes on average only a few posts. This means that a user appearing in the training set is less likely to be present in the test set as well. Even in this case,  user embeddings are still effectively learned thanks to the generalizability of the  \textit{doc2vec} model.

\subsubsection{Integration}
The outputs $X_F$ and $U_F$ of the text and user modules are concatenated to achieve a high-level  representation characterizing the propagation dynamics of news. 
The concatenated vector is then fed to a fully connected layer, performing linear and softmax transformations to obtain the final prediction. We use the cross entropy loss  for binary classification with labels \{\textit{rumour, non-rumour}\} as objective function, and we minimize it using the Adam algorithm~\cite{kingma2014adam}.

% et the output of the DRRD model and the ground-truth label for the $i$-th event be $\tilde{L}^{(i)}$ and $L^{(i)}$, respectively, the objective function is given by
% 
% \begin{equation}
% \mathcal{L} = -\frac{1}{N}\sum^N_{i=1} \sum^D_{\mathcal{C}=1} L^{(i)}_j % \log(\tilde{L}^{(i)}_j) \ ,
% \end{equation}
% where $N$ is the number of events and $D$ is the number of considered classes. 
% We optimize $\mathcal{L}$ using the Adam algorithm~\cite{kingma2014adam}. 

\begin{table}[t]
        \caption{\label{tab:datasets}The description of the Weibo and Twitter
datasets.}
        \centering
        \begin{tabular}{c | c | c | c }
            \hline
            \hline
                & Weibo & Twitter &\hspace{-5pt} Twitter (incomplete)\hspace{-5pt}
                \\
                \hline
                \hspace{-5pt}Num. users \hspace{-5pt} & 2.819.338 & 233.719
& 210.838 \\
               \hspace{-5pt} Num. events\hspace{-5pt} & 4664 & 992 & 991
\\
                \hspace{-5pt}Num. posts\hspace{-5pt} & 3.752.459 & 592.391
& 510.147 \\
                \hspace{-5pt}Num. rumours \hspace{-5pt}& 2313 & 498 & 498
\\
               \hspace{-5pt} Num. non-rumours\hspace{-4pt}  & 2351 & 494
& 493 \\
                \hline
                \hline
        \end{tabular}
\end{table}

\section{Experiments}\label{sec:experiments}

\subsection{Datasets}\label{sec:dataset}
We employed two real-world datasets to evaluate the proposed model, which are collected from Weibo~\cite{ma2016detecting} and Twitter~\cite{castillo2011information, ma2015detect, ma2016detecting}, respectively. 
Table~\ref{tab:datasets} gives the description of these datasets. 
Only the IDs of relevant posts and the labels for each event are provided in each dataset, which means that one needs to crawl the data from the Weibo and Twitter application programming interfaces (APIs).  
The posts in the Weibo dataset can be retrieved completely, while in the Twitter dataset, many tweets were removed, thus it cannot be retrieved completely via the Twitter API. 
According to our calculation, the number of missing tweets is about $13.8\%$ of the original number reported in~\cite{ma2016detecting}. Our experiments are conducted on the Weibo and the \textit{incomplete Twitter} datasets. In what follows, when we mention the Twitter dataset we refer to the \textit{incomplete Twitter} dataset. 

\subsection{Experimental Setting}
\newcommand\Tstrut{\rule{0pt}{2.0ex}}         % = `top' strut
\newcommand\Bstrut{\rule[-0.5ex]{0pt}{0pt}}   % = `bottom' strut
\begin{table*}[t]
    \centering
    \caption{Extended rumour detection performance of the DRRD model in comparison with baseline models (R:Rumour, N:Non-Rumour)}
    \label{tab:res_comp}
    \begin{tabular}{c c c c c c c c c c}
        \hline
        \hline
        \Tstrut\multirow{2}{*}{Model} & \multirow{2}{*}{Class} & \multicolumn{4}{c}{Weibo} & \multicolumn{4}{c}{Twitter} \\
         & & Accuracy &Precision& Recall& $F_1$ & Accuracy &Precision &Recall& $F_1$ \Bstrut\\
         \hline
         \Tstrut\multirow{2}{*}{SVM-RBF} & R &  \multirow{2}{*}{0.818} & 0.822 & 0.812 & 0.817 &  \multirow{2}{*}{0.715} & 0.698 & 0.809 & 0.749 \\
          & N &  & 0.815 & 0.824 & 0.819 &  & 0.741 & 0.610 & 0.669\\
         \Tstrut\multirow{2}{*}{DTC} & R &  \multirow{2}{*}{0.831} & 0.847 & 0.815 & 0.831 &  \multirow{2}{*}{0.718} & 0.721 & 0.711 & 0.716 \\
          & N &  & 0.815 & 0.847 & 0.830 &  & 0.715 & 0.725 & 0.720\\
         \Tstrut\multirow{2}{*}{RFC} & R &  \multirow{2}{*}{0.849} & 0.786 & 0.959 & 0.864 &  \multirow{2}{*}{0.728} & 0.742 & 0.737 & 0.740 \\
          & N &  & 0.947 & 0.739 & 0.830 &  & 0.713 & 0.718 & 0.716\\
         \Tstrut\multirow{2}{*}{SVM-TS} & R &  \multirow{2}{*}{0.857} & 0.878 & 0.830 & 0.857 &  \multirow{2}{*}{0.745} & 0.707 & 0.864 & 0.778 \\
          & N &  & 0.947 & 0.739 & 0.830 &  & 0.809 & 0.618 & 0.701\\
         \Tstrut\multirow{2}{*}{GRU-2} & R &  \multirow{2}{*}{0.910} & 0.876 & 0.956 & 0.914 &  \multirow{2}{*}{0.757} & 0.732 & 0.815 & 0.771 \\
          & N &  & 0.952 & 0.864 & 0.906 &  & 0.788 & 0.698 & 0.771\\
          
         \Tstrut\multirow{2}{*}{CAMI} & R &  \multirow{2}{*}{0.933} & 0.921 & 0.945 & 0.933 &  \multirow{2}{*}{0.777} & 0.744 & \textbf{0.848} & 0.793 \\
          & N &  & 0.945 & 0.921 & 0.932 &  & \textbf{0.820} & 0.705 & 0.758 \Bstrut\\
        
                \Tstrut\multirow{2}{*}{CSI} & R &  \multirow{2}{*}{0.932} & 0.938 & 0.924 & 0.931 &  \multirow{2}{*}{0.787} & 0.755 & 0.854 & 0.802 \\
          & N &  & 0.926 & 0.94 & 0.933 &  & 0.828 & 0.719 & 0.77 \Bstrut\\

         \hline

\Tstrut\multirow{2}{*}{SRRD} & R &  \multirow{2}{*}{0.949} & 0.953 & 0.944 & 0.949 &  \multirow{2}{*}{0.748} & 0.764 & 0.723 & 0.743 \\
          & N &  & 0.946 & 0.955 & 0.950 &  & 0.732 & 0.773 & 0.752 \Bstrut\\
          
          \Tstrut\multirow{2}{*}{DRRD} & R &  \multirow{2}{*}{\textbf{0.968}} & 0.959 & \textbf{0.979} & \textbf{0.969} &  \multirow{2}{*}{\textbf{0.806}} & 0.817 & 0.795 & 0.804 \\
          & N &  & \textbf{0.978} & 0.958 & 0.968 &  & 0.798 & 0.804 & \textbf{0.809} \Bstrut\\          
          
         \hline
                \hline
        
    \end{tabular}
\end{table*}

%\begin{figure}[h!]
%\centering
%\begin{subfigure}{\linewidth}
%  \centering
%  \includegraphics[width=0.7\textwidth]{Images/weibo_early.png}
%  \caption{Weibo}
%  \label{fig:sub1}
%\end{subfigure}
%\vspace{-3mm}
%\begin{subfigure}{\linewidth}
%  \centering
%  \includegraphics[width=0.7\textwidth]{Images/twitter_early.png}
%  \caption{Twitter}
%  \label{fig:sub2}
%\end{subfigure}
%\caption{Early identification performance comparison}
%\label{fig:test}
%\end{figure}

For the \textit{doc2vec} model, we employ the Distributed Bag-of-Word (DBOW) version with $d_v = 100$ dimensions for both the text and user embeddings. 
In the RNN network, we set the number of hidden units to $d_f = 128$ for both hidden layers. Similarly, the final fully connected layer has $128$ hidden units. In all layers,  
we use the \texttt{tanh} as activation function. To avoid overfitting, we use dropout regularization~\cite{srivastava2014dropout} for the RNN and the final fully connected layer. We empirically choose a dropout rate of $0.6$. 
Our model is implemented using Tensorflow. 

%\begin{figure}[h]
%\centering
%\includegraphics[width=\linewidth]{Images/early.png}
%\caption{Early detection performance of baselines and our method on the Twitter (left) and Weibo (right) datasets.}
%\label{fig:early_detection}
%\end{figure}

In order to evaluate the performance of the proposed model, we conduct experiments using two settings. 
In the first setting, all the posts in the entire time-span of the given dataset are considered. We call it the \textit{extended detection} setting. 
In the second setting, we consider only the posts appeared within specific deadlines; this setting is referred to as \textit{early detection}. 
In both settings, we adhere to the data splitting that is considered in previous studies~\cite{ma2016detecting, yu2017convolutional}. Namely, for each dataset, we hold a random set of 10\% of events for model fine-tuning. 
The rest of the events are split with a 3:1 ratio for training and testing, respectively, leading to a 4-fold cross validation scheme. 
Similar to~\cite{ma2016detecting, yu2017convolutional}, we compare our method against the following  schemes: 1)~\textbf{SVM-RBF}~\cite{yang2012automatic}, 2)~\textbf{DTC}~\cite{castillo2011information}, 3)~\textbf{RFC}~\cite{kwon2013prominent}, 4)~\textbf{SVM-TS}~\cite{ma2015detect}, 5)~\textbf{GRU-2}~\cite{ma2016detecting}, 6)~\textbf{CAMI}~\cite{yu2017convolutional} and 7)~\textbf{CSI}~\cite{ruchansky2017csi}.  
The results of the first six methods are taken from~\cite{yu2017convolutional, ma2015detect}, whereas those of  the CSI model~\cite{ruchansky2017csi} are obtained by our implementation. This is  because the evaluation in~\cite{ruchansky2017csi}  considers a different dataset splitting strategy. 
Furthermore, in order to validate the capacity of our DRRD model in learning
user representations, we replace the proposed user module with the  SVD-based module presented in~\cite{ruchansky2017csi}. 
We refer to this modified DRRD model as the SRRD model (SVD-based RNN rumour detection). 
We assess the performance of the considered models in terms of the  accuracy, precision, recall and F1-score metrics.

\subsection{Extended Rumour Detection Results}\label{sec:extended_detection}
\begin{figure}[t]
\centering
\includegraphics[width=\linewidth]{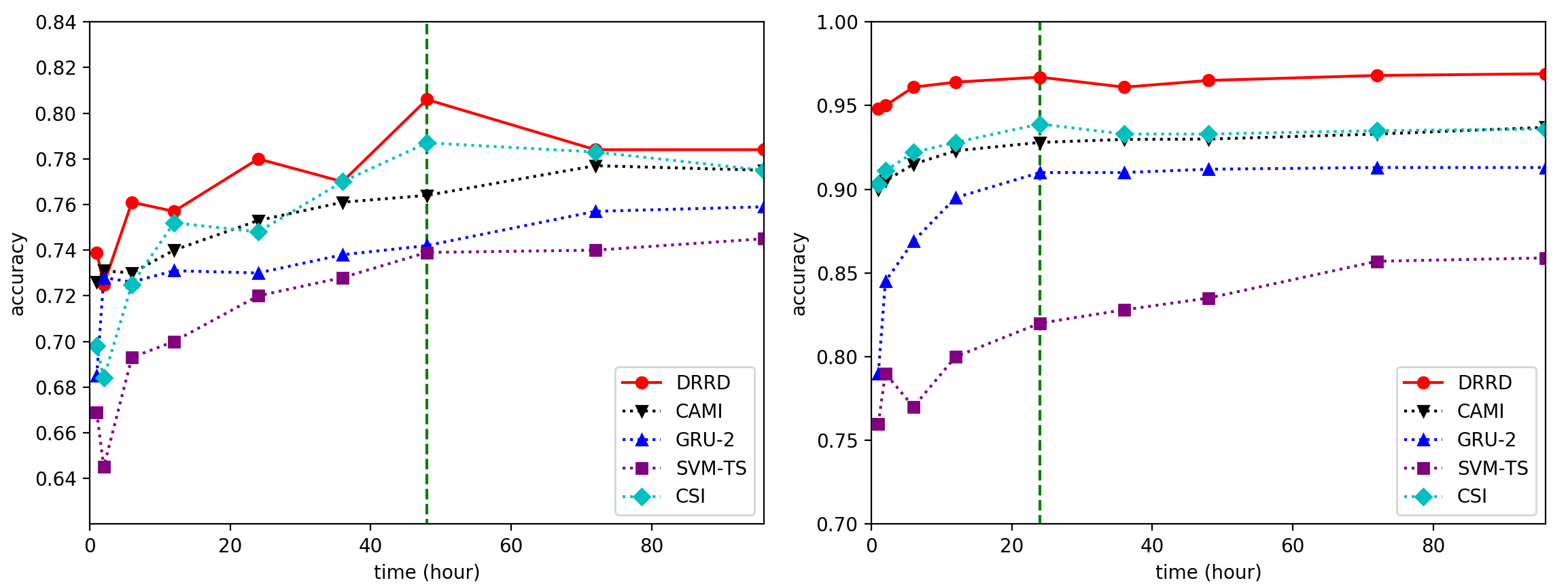}
\caption{Early detection performance of baselines and our method on the incomplete Twitter (left) and Weibo (right) datasets.}
\label{fig:early_detection}
\end{figure}

The results for the proposed model (both the DRRD and the SRRD versions) and the baselines are reported in Table~\ref{tab:res_comp}. 
The CAMI and CSI models, which are deep-learning-based models, achieve good performance;
nevertheless, the proposed model delivers the best performance for both datasets. Specifically, our model yields the best detection results in terms of accuracy, precision, recall and F1-score on the Weibo dataset. On the Twitter dataset, our model achieves comparable results with other models in terms of  the precision and recall metrics, and the best results in terms of the  accuracy and F1-score metrics.

Furthermore, the results in Table~\ref{tab:res_comp} corroborate the superior performance of the proposed user module in learning user representation in comparison with the SVD-based approach (see results obtained with the SRNN version). Specifically, using the proposed user module improves the accuracy by more than 2\% compared to the SVD-based counterpart on both the Weibo and Twitter datasets. It also leads to better performance  in terms of the precision, recall and F1-score metrics on both datasets.  

\subsection{Early Rumour Detection Results}\label{sec:early_detection}

Figure~\ref{fig:early_detection} shows the accuracy of the DRRD model and the baseline models on the Weibo and Twitter datasets for the early detection setting. 
On the Weibo dataset, the proposed model outperforms the other models at all considered deadlines and the best performance of the DRRD is achieved when $T = 24$ h. 
Although we observe some fluctuations in the performance on the Twitter dataset, the DRRD model still outperforms the other models at most of the deadlines, with the best performance obtained when $T=48$ h.

The reasons
that explain why the proposed model can detect rumours effectively within
the very first hours after an event starts circulating on social media are as follows. 
Firstly, as illustrated in Fig.~\ref{fig:post_pattern}, most of the social media posts are made during the  first few hours following the publication of an article. Secondly, the variation in the number of posts is more pronounced during these first hours. The higher the variation in the number of posts, the more information it reveals about the propagation process. Alternatively, one may notice that the performance of DRRD slightly decreases when more data is available (e.g., $T = 84$). This is because the propagation patterns of rumours and genuine news tend to be similar over time.
%\vspace{-5mm}

\section{ Conclusion}\label{sec:conclusion}
Misleading information is an important issue nowadays with  serious consequences. There have been many studies  addressing this problem, however, detecting this kind of disinformation effectively and timely still remains a challenging task. In this work, we presented a deep neural-network-based model capable of detecting rumours via  learning propagation dynamics and user representations. 
The proposed model was shown to achieve superior results compared to various state-of-the-art models on two benchmark datasets. 
%The future work will focus on exploiting the content of rumours, including their enclosed media such as images and videos, and combining the content-based discriminating features with the propagation pattern. 

%\vfill
%\pagebreak

%\vfill\pagebreak
%\newpage

\bibliographystyle{IEEEbib}
\bibliography{refs}

\begin{thebibliography}{10}

\bibitem{zubiaga2018detection}
A.~Zubiaga, A.~Aker, K.~Bontcheva, M.~Liakata, and R.~Procter,
\newblock ``Detection and resolution of rumours in social media: A survey,''
\newblock {\em ACM Computing Surveys}, vol. 51, pp. 32, 2018.

\bibitem{shu2017fake}
K.~Shu, A.~Sliva, S.~Wang, J.~Tang, and H.~Liu,
\newblock ``Fake news detection on social media: A data mining perspective,''
\newblock {\em ACM SIGKDD Explorations Newsletter}, vol. 19, pp. 22--36, 2017.

\bibitem{ciampaglia2015computational}
G.~L. Ciampaglia, P.~Shiralkar, L.~M. Rocha, J.~Bollen, F.~Menczer, and
  A.~Flammini,
\newblock ``Computational fact checking from knowledge networks,''
\newblock {\em PloS one}, vol. 10, 2015.

\bibitem{rubin2015truth}
V.~L. Rubin and T.~Lukoianova,
\newblock ``Truth and deception at the rhetorical structure level,''
\newblock {\em JASIST}, vol. 66, pp. 905--917, 2015.

\bibitem{duc2019naccle}
D.~M. Nguyen, T.~H. Do, R.~Calderbank, and N.~Deligiannis,
\newblock ``Fake news detection using deep markov random fields,''
\newblock in {\em NAACL}, 2019, pp. 1--10.

\bibitem{jin2016news}
Z.~Jin, J.~Cao, Y.~Zhang, and J.~Luo,
\newblock ``News verification by exploiting conflicting social viewpoints in
  microblogs,''
\newblock in {\em AAAI}, 2016, pp. 2972--2978.

\bibitem{wu2015false}
K.~Wu, S.~Yang, and K.~Q. Zhu,
\newblock ``False rumors detection on sina weibo by propagation structures,''
\newblock in {\em ICDE}, 2015, pp. 651--662.

\bibitem{castillo2011information}
C.~Castillo, M.~Mendoza, and B.~Poblete,
\newblock ``Information credibility on twitter,''
\newblock in {\em WWW}, 2011, pp. 675--684.

\bibitem{ma2016detecting}
J.~Ma, W.~Gao, P.~Mitra, S.~Kwon, B.~Jansen, K.~F. Wong, and M.~Cha,
\newblock ``Detecting rumors from microblogs with recurrent neural networks,''
\newblock in {\em IJCAI}, 2016, pp. 3818--3824.

\bibitem{ma2015detect}
J.~Ma, W.~Gao, Z.~Wei, Y.~Lu, and K.~F. Wong,
\newblock ``Detect rumors using time series of social context information on
  microblogging websites,''
\newblock in {\em CIKM}, 2015, pp. 1751--1754.

\bibitem{magdy2010web}
A.~Magdy and N.~Wanas,
\newblock ``Web-based statistical fact checking of textual documents,''
\newblock in {\em Int. workshop on Search and mining user-generated contents},
  2010, pp. 103--110.

\bibitem{mukherjee2013fake}
A.~Mukherjee, V.~Venkataraman, B.~Liu, and N.~Glance,
\newblock ``Fake review detection: Classification and analysis of real and
  pseudo reviews,''
\newblock {\em UIC-CS-03-2013. Technical Report}, 2013.

\bibitem{rashkin2017truth}
H.~Rashkin, E.~Choi, J.~Y. Jang, S.~Volkova, and Y.~Choi,
\newblock ``Truth of varying shades: On political fact-checking and fake
  news,''
\newblock in {\em EMNLP}, 2017.

\bibitem{potthast2017stylometric}
M.~Potthast, J.~Kiesel, K.~Reinartz, J.~Bevendorff, and B.~Stein,
\newblock ``A stylometric inquiry into hyperpartisan and fake news,''
\newblock {\em arXiv:1702.05638}, 2017.

\bibitem{yang2018ti}
Y.~Yang, L.~Zheng, J.~Zhang, Q.~Cui, Z.~Li, and P.~S. Yu,
\newblock ``Ti-cnn: Convolutional neural networks for fake news detection,''
\newblock {\em arXiv:1806.00749}, 2018.

\bibitem{wang2017liar}
W.~Y. Wang,
\newblock ``liar, liar pants on fire: A new benchmark dataset for fake news
  detection,''
\newblock {\em arXiv:1705.00648}, 2017.

\bibitem{zhang2018fake}
J.~Zhang, L.~Cui, Y.~Fu, and F.~B. Gouza,
\newblock ``Fake news detection with deep diffusive network model,''
\newblock {\em arXiv:1805.08751}, 2018.

\bibitem{tacchini2017some}
E.~Tacchini, G.~Ballarin, M.~L.~D. Vedova, S.~Moret, and L.~Alfaro,
\newblock ``Some like it hoax: Automated fake news detection in social
  networks,''
\newblock {\em arXiv:1704.07506}, 2017.

\bibitem{zubiaga2016learning}
A.~Zubiaga, M.~Liakata, and R.~Procter,
\newblock ``Learning reporting dynamics during breaking news for rumour
  detection in social media,''
\newblock {\em arXiv:1610.07363}, 2016.

\bibitem{jin2014news}
Z.~Jin, J.~Cao, Y.~G. Jiang, and Y.~Zhang,
\newblock ``News credibility evaluation on microblog with a hierarchical
  propagation model,''
\newblock in {\em IEEE ICDM}, 2014, pp. 230--239.

\bibitem{chen2017call}
T.~Chen, L.~Wu, X.~Li, J.~Zhang, H.~Yin, and Y.~Wang,
\newblock ``Call attention to rumors: Deep attention based recurrent neural
  networks for early rumor detection,''
\newblock {\em arXiv:1704.05973}, 2017.

\bibitem{kochkina2018all}
E.~Kochkina, M.~Liakata, and A.~Zubiaga,
\newblock ``All-in-one: Multi-task learning for rumour verification,''
\newblock {\em arXiv:1806.03713}, 2018.

\bibitem{yu2017convolutional}
F.~Yu, Q.~Liu, S.~Wu, L.~Wang, and T.~Tan,
\newblock ``A convolutional approach for misinformation identification,''
\newblock in {\em IJCAI}, 2017, pp. 3901--3907.

\bibitem{qian2018neural}
F.~Qian, C.~Gong, K.~Sharma, and Y.~Liu,
\newblock ``Neural user response generator: Fake news detection with collective
  user intelligence.,''
\newblock in {\em IJCAI}, 2018, pp. 3834--3840.

\bibitem{liu2018earlydetect}
Y.~Liu and Y.~B. Wu,
\newblock ``Early detection of fake news on social media through propagation
  path classification with recurrent and convolutional networks,''
\newblock in {\em AAAI}, 2018.

\bibitem{ruchansky2017csi}
N.~Ruchansky, S.~Seo, and Y.~Liu,
\newblock ``Csi: A hybrid deep model for fake news detection,''
\newblock in {\em CIKM}, 2017, pp. 797--806.

\bibitem{le2014distributed}
Q.~Le and T.~Mikolov,
\newblock ``Distributed representations of sentences and documents,''
\newblock in {\em ICML}, 2014, pp. 1188--1196.

\bibitem{do2017multiview}
Tien~Huu Do, Duc~Minh Nguyen, Evaggelia Tsiligianni, Bruno Cornelis, and Nikos
  Deligiannis,
\newblock ``Multiview deep learning for predicting twitter users' location,''
\newblock {\em arXiv:1712.08091}, 2017.

\bibitem{lipton2015critical}
Z.~Lipton, J.~Berkowitz, and C.~Elkan,
\newblock ``A critical review of recurrent neural networks for sequence
  learning,''
\newblock {\em arXiv:1506.00019}, 2015.

\bibitem{gru2014learning}
K.~Cho, B.~V. Merri{\"e}nboer, C.~Gulcehre, D.~Bahdanau, F.~Bougares,
  H.~Schwenk, and Y.~Bengio,
\newblock ``Learning phrase representations using rnn encoder-decoder for
  statistical machine translation,''
\newblock {\em arXiv:1406.1078}, 2014.

\bibitem{lstm1997long}
S.~Hochreiter and J.~Schmidhuber,
\newblock ``Long short-term memory,''
\newblock {\em Neural computation}, vol. 9, pp. 1735--1780, 1997.

\bibitem{doc2014distributed}
Q.~Le and T.~Mikolov,
\newblock ``Distributed representations of sentences and documents,''
\newblock in {\em ICML}, 2014, pp. 1188--1196.

\bibitem{kingma2014adam}
D.~Kingma and J.~Ba,
\newblock ``Adam: A method for stochastic optimization,''
\newblock {\em arXiv:1412.6980}, 2014.

\bibitem{srivastava2014dropout}
N.~Srivastava, G.~Hinton, A.~Krizhevsky, I.~Sutskever, and R.~Salakhutdinov,
\newblock ``Dropout: a simple way to prevent neural networks from
  overfitting,''
\newblock {\em JMLR}, vol. 15, pp. 1929--1958, 2014.

\bibitem{yang2012automatic}
F.~Yang, Y.~Liu, X.~Yu, and M.~Yang,
\newblock ``Automatic detection of rumor on sina weibo,''
\newblock in {\em ACM SIGKDD Workshop on Mining Data Semantics}, 2012, p.~13.

\bibitem{kwon2013prominent}
S.~Kwon, M.~Cha, K.~Jung, W.~Chen, et~al.,
\newblock ``Prominent features of rumor propagation in online social media,''
\newblock in {\em IEEE ICDM}, 2013, pp. 1103--1108.

\end{thebibliography}

\end{document}